# SIMPLE CRACKING OF (NOISE-BASED) DYNAMIC WATERMARKING IN SMART GRIDS

MEHMET YILDIRIM [1], NASIR KENARANGUI [1], ROBERT BALOG [1], LASZLO B. KISH [1,2,†], CHANAN SINGH [1]

[1]*Department of Electrical and Computer Engineering, Texas A&M University, TAMUS 3128, College Station, TX, USA*

[2]*Óbuda University, Budapest, Bécsi út 96/B, Budapest, H-1034, Hungary*

Previous research employing a conceptual approach with a digital twin has demonstrated that (noise-based) dynamic watermarking is incapable of providing unconditional security in smart electrical grid systems. However, the implementation of digital twins can be prohibitively costly or infeasible due to limited available data on critical infrastructure. In this study, we first analyze the spectral properties of dynamic watermarking and its associated protocol. Subsequently, we present a straightforward attack inspired by the digital twin method, which extracts and utilizes the grid noises and completely breaches the security of dynamic watermarking without requiring knowledge of the private watermarking signal. The attacker can fully expose the grid while evading detection by the controller. Our findings indicate that in the absence of secure and authenticated communications, dynamic watermarking offers neither conditional nor unconditional security. Conversely, when communication lines, sensors, and communicators are equipped with tamper-resistant and secure/authenticated links, dynamic watermarking becomes redundant for grid security.

## 1. Introduction

Recently, we proved [1] that the dynamic watermarking (DW) method [2-5] does not provide unconditional security for smart electrical grid systems. The proof involved a digital twin utilized by an (unlimited) adversary (troublemaker, Trudy) to extract the DW component from the original sensor signal.

While the aforementioned proof has a theoretical role, it is essential to acknowledge that digital twins [6] are not only non-perfect but also expensive in practical situations. In this paper, we demonstrate that a similarly successful attack can be achieved through a much simpler, and less expensive, approach. One of the key conclusions is that Trudy does not need to be an unlimited adversary to crack the DW scheme. We present a simple hardware scheme and protocol capable of defeating the system, implying that the DW scheme is neither unconditionally nor conditionally secure against this attack.

*Section Overview*: Section 2 presents a simple example of the DW protocol, analyzing its spectral properties and the implications for some necessary operational conditions. Section 3 circumvents the purported DW security with a new and simple attack protocol inspired by the previous digital twin method [1].

---

[†] Corresponding author. Honorary faculty at Óbuda Univ.

## 2. The dynamic watermarking scheme and its spectral analysis

*2.1 The dynamically watermarked ideal grid*

The goal of the DW protocol [2-5] is to secure smart power grids without utilizing secure communications for the sensor signals, see Figure 1. Within his private space of the control center, Conrad controls operational parameters (voltage amplitude, frequency and phase) of the smart grid. He injects a secret dynamical watermark, a low-frequency, Gaussian noise $N_w(t)$, into the error voltage channel of the control unit. That yields a small random modulation of the envelope of the sinusoidal line voltage. Without loss of generality, we suppose a 60Hz nominal grid frequency (note small frequency fluctuations around 60Hz do not impact this analysis).

In the grid at *idealized* conditions, assuming zero voltage drop on line impedance and a balanced system with no parasitic noises, the representative line voltage is given as:

$$V_L(t) = a_0 \sin(2\pi f_g t + \varphi_0) = a_0 \sin(2\pi 60 t + \varphi_0) \; . \tag{1}$$

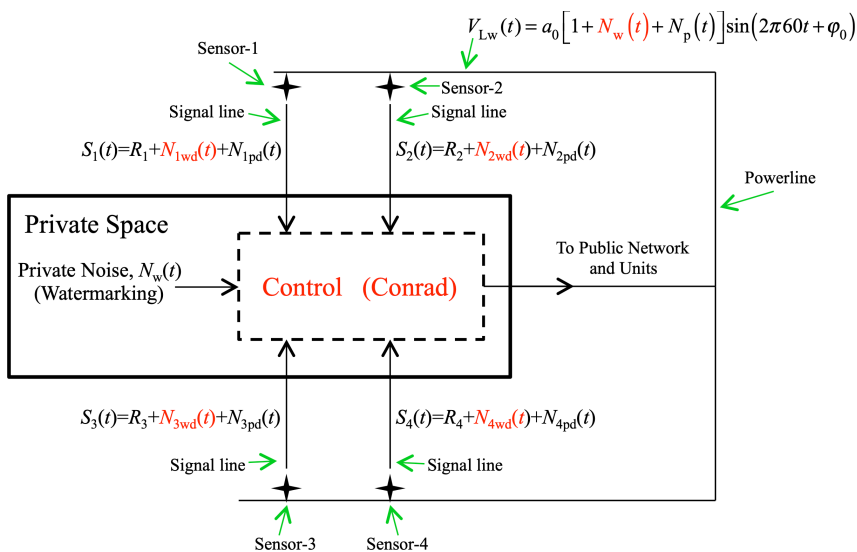

Figure 1. Illustration of the dynamically watermarked power grid with voltage sensors, where $N_w(t)$ represents the watermarking Gaussian noise injected by Conrad into the control unit. This noise will cause a small random amplitude modulation of the envelope (peak voltage) $a_0$. At practical conditions, there are also other "parasitic" deviations $N_p(t)$ that originate from the normal operation of the grid, see Section 2.4. The practical peak voltage may differ from the regular one due to inaccuracies but that is also represented by the $N_p(t)$ function. In the ideal cases, when there is no attack, the sensors detect the regular (nominal) peak voltage $R=a_0$ and the two noise components, $N_{wd}$ and $N_{pd}$ superimposed on them, respectively. These quantities are indexed according to the sensor.

In the same idealized grid, the line voltage dynamically watermarked by Conrad, $V_{Lw}(t)$, contains a small random amplitude modulation due to $N_w(t)$:

$$V_{Lw}(t) = a(t)\sin(2\pi 60 t + \varphi_0) = a_0 \big[1 + N_w(t)\big]\sin(2\pi 60 t + \varphi_0) \; , \tag{2}$$

where the dynamically watermarked envelope is:

$$E_w(t) = a_0\left[1 + N_w(t)\right] . \tag{3}$$

The modulation must be sufficiently small that it should not interfere with the normal grid operations (such as under- or over-voltage protection), thus the mean-square modulation must satisfy:

$$\left\langle N_w^2(t) \right\rangle \ll 1 . \tag{4}$$

*2.2 Spectral properties of the dynamically watermarked grid*

From Equation 2, it follows that the instantaneous grid voltage is the sum of the original sinusoidal component at the grid operating frequency ($f_g$) and the noise component, which is the modulation product:

$$V_{Lw}(t) = E_w(t)\sin(2\pi 60 t + \varphi_0) = a_0 \sin(2\pi 60 t + \varphi_0) + a_0 N_w(t)\sin(2\pi 60 t + \varphi_0) . \tag{5}$$

Suppose that bandwidth of the DW noise is *B*. According to the elements of analog modulation (AM), the frequency band of the modulation product is symmetrically located around the 60Hz "carrier frequency" and it has 2*B* as the bandwidth.

The power density spectra *S(f)* are illustrated in Figure 2. In the frequency range [0,*B*] the band-limited white spectrum of Conrad's DW noise $N_w(t)$ is shown. The Dirac pulse at the grid frequency (60Hz) and the band-limited white noise spectrum with 2B bandwidth around it represent the power density spectrum of the line voltage. The scales are arbitrary; the figure serves only as an illustration of the frequency distribution.

Why is this important? Power grids are designed to operate over a narrow frequency range. The allowable range of grid frequency in USA is 59.7Hz to 60.3Hz. Thus even though the modulation is shallow, it is advisable to stay in the frequency range where the frequency components of the noise modulation stay within the bandwidth of the allowable grid operating frequency. Furthermore, the control system of the generators has various time constants [7] that limit the frequency bandwidth *B* of the DW noise typically to 0.3 Hz or less [7].

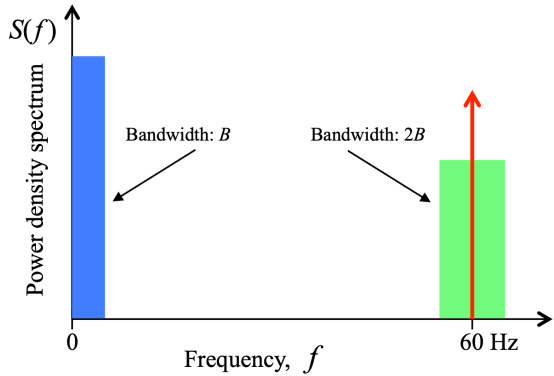

Figure 2. The spectra of the watermarking noise (left) and the watermarked line voltage (right). The Dirac pulse is the original spectrum of the line voltage before watermarking. The spectrum of the modulation product flatly spreads over the frequency range of [60-*B*, 60+*B*].

In conclusion, it is advisable to have $B \leq 0.3\text{Hz}$.

*2.3 Voltage sensor operation*

Conrad needs *amplitude demodulation* to determine the integrity of the envelope $E_w(t)$. For example, synchronous detection is multiplying with a phase-synchronized sinusoidal signal $\sin(2\pi 60 t + \varphi_0)$, such as obtained from the use of a phase-locked loop. That results in the following formula:

$$V_{\text{Lw}}(t)\sin(2\pi 60 t + \varphi_0) = a_0 \frac{1-\cos(2\pi 120 t + 2\varphi_0)}{2} + a_0 N_w(t)\frac{1-\cos(2\pi 120 t + 2\varphi_0)}{2} \quad . \quad (6)$$

The relevant power density spectra *S(f)* are illustrated in Figure 3. In the frequency range [0,*B*] the restored band-limited white spectrum of Conrad's DW noise $N_w(t)$ and a Dirac pulse at zero indicating a DC component are shown. The Dirac pulse at 120Hz and the band-limited white noise spectrum with 2*B* bandwidth around it represent the power density spectrum of the non-desired components of the demodulation. The scales are arbitrary; the figure serves only as an illustration of the frequency distribution.

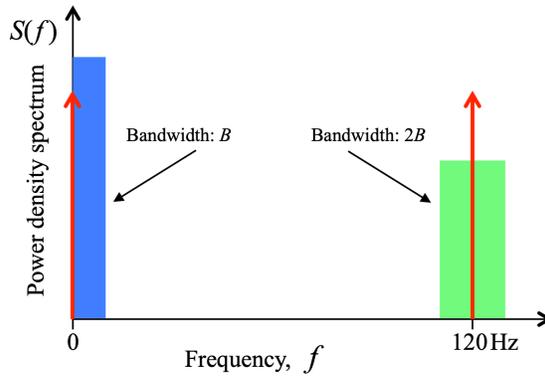

Fig. 3. The spectra during the demodulation process. The Dirac pulses are at DC and 120 Hz. To obtain the original watermarking noise at the low-frequency end, the voltage must be filtered or time averaged for 1/120 second and the DC component must be subtracted.

Thus the envelope can be extracted by a finite-time time-averaging over the period of the second harmonic:

$$\langle 2V_{\text{Lw}}(t)\sin(2\pi 60 t + \varphi_0)\rangle_\tau \approx a_0[1+N_{\text{wd}}(t)] = E_{\text{wd}}(t) \quad (7)$$

where $E_{\text{wd}}(t)$ and $N_{\text{wd}}(t)$ are the detected envelope and watermarking signal, respectively; the averaging time is $\tau = \frac{1}{2f} = \frac{1}{120}$; and the approximation is valid for $B \ll 60\text{Hz}$.

This process allows monitoring the envelope and the watermarking noise $N_{\text{wd}}(t)$ with a sampling frequency of $2f = 120 Hz$. Alternatively, a low-pass filter can be applied to remove the frequency components above $120 - B$, which will however imply a somewhat reduced sampling rate.

*2.4 Conrad's security check*

Conrad is receiving $E_w(t)$ from the sensors and checking the existence of $N_{wd}(t)$ in $E_w(t)$. We suppose that time delays are negligible. Theoretically, such a test could be done with a direct comparison of the original $N_w(t)$ and the detected $N_{wd}(t)$ signals. However, the practically detected envelope $E_{wdp}(t)$ can typically contain other noises and inaccuracies steaming from the random events during the operation of the grid, thus the practical watermarked envelope consists additional deviations $N_p(t)$, too, see Figure 1. Thus after the measurement by the sensor, such a practical envelope will be:

$$E_{wdp}(t) = a_0 \left[ 1 + N_{wd}(t) + N_{pd}(t) \right] . \qquad (8)$$

Because $N_{pd}(t)$ contains random components, too, Conrad require statistical tools to test for the presence of the watermarking noise. For example, he can test the variance of $E_{wdp}(t)$ as it was proposed and demonstrated in the original DW papers [2-5].

Alternatively, Conrad can detect the existence of $N_{wd}(t)$ by cross-correlation technique:

$$D_w = \frac{\left\langle E_{wdp}(t) N_w(t) \right\rangle_{T_0}}{a_0} , \qquad (9)$$

where $T_0$ is the averaging time. Ideally, for infinite $T_0$, $D_w$ is either 1 or 0. When it is 0, Conrad interprets the situation so that an attack is going on. Due to the small-noise condition posed by Equation 1, in practical situations, with finite $T_0$, the result will be between 0 and 1 and a proper threshold must be chosen for the protocol to interpret an ongoing attack vs. an attack-free situation.

**3. Cracking the DW scheme by simple means**

The proposed attack is the simplification and improvement of the digital twin based attack [1]. See Figure 4 for an example of attacking a voltage sensor, Sensor-1. Trudy breaks the communication line of Sensor-1 toward Conrad and injects a fake sensor signal that contains the faked voltage data and the expected watermarking and parasitic noise signals.

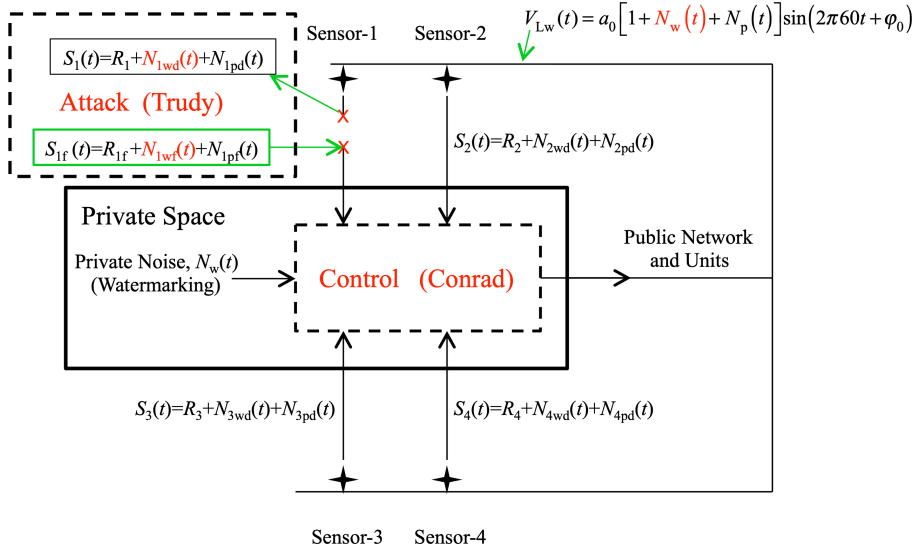

Figure 4. Simple attack against the DW scheme. A fake signal of Sensor-1 containing the extracted watermarking noise is synthesized by Trudy. The nominal value $a_0$ of the envelope is subtracted from the sensor signal. Only the two noise components remain. The properly scaled noise components will be the fake noises that are added to Trudy's fake sensor signal. In [1] the same process was carried out except that the "nominal" value to subtract was generated by a digital twin, which is a difficult solution.

The generic form of the sensor signal $S(t)$ is given as:

$$S(t) = R + N_{wd}(t) + N_{pd}(t) = a_0\left[1 + N_{wd}(t) + N_{pd}(t)\right] \qquad (10)$$

Based on Equation 8 and 10, Trudy can extract the noise component $E_{wdpN}(t)$ of the detected envelope $E_{wdp}(t)$ of the sensor signal by subtracting the nominal peak voltage $a_0$:

$$E_{wdpN}(t) = E_{wdp}(t) - a_0 = a_0\left[1 + N_{wd}(t) + N_{pd}(t)\right] - a_0 = a_0\left[N_{wd}(t) + N_{pd}(t)\right]. \qquad (11)$$

By having the critical watermarking component, Trudy can synthesize a fake sensor signal $S_f(t)$:

$$S_f(t) = R_f + N_{wdf}(t) + N_{pdf}(t) \qquad (12)$$

The reason for the "fake" subscript at each term in the equation is that she needs to properly scale the fake (extracted) watermarking noise and fake (extracted) parasitic noise to match the changes she introduced with the fake sensor signal. In general:

$$S_f(t) = R_f + N_{wdf}(t) + N_{pdf}(t) = \alpha R + \beta N_{wd}(t) + \gamma N_{pdf}(t), \qquad (13)$$

where the values of $\alpha$ $\beta$ and $\gamma$ must be chosen according to the imitated failure.

For instance, if Trudy aims to emulate an equipment malfunction or failure that leads to a reduced grid voltage, such as a transformer fault, she would scale down the nominal voltage and the corresponding terms proportionally to mimic the voltage reduction effects associated with the simulated fault condition:

$$\alpha = \beta = \gamma < 1 \ . \tag{14}$$

Another scenario involves a hybrid power system where the total power supply comprises a dynamically watermarked main grid component by Conrad and a non-watermarked renewable energy source, such as solar power. In such cases, the scaling factors applied to each power component would differ to account for the presence of watermarking in only one of the sources:

$$\alpha \neq \beta \neq \gamma \ . \tag{15}$$

If an entity, such as Conrad, intends to dynamically watermark the renewable energy component, for instance, solar power, they must establish secure communication channels to transmit their private noise sequence ($N_w$) to the inverters at the solar plants. However, implementing secure communication infrastructures across the grid would render DW redundant, as the secure channels could also guarantee the integrity of sensor signals and provide tamper-resistance capabilities, thereby negating the need for watermarking as a security measure.

## 4. Illustrations of the basic signal shapes

Below, a few illustrative graphs are shown about the signal shapes discussed above. Figure 5 illustrates the amplitude of a power line of 100kV rms value.

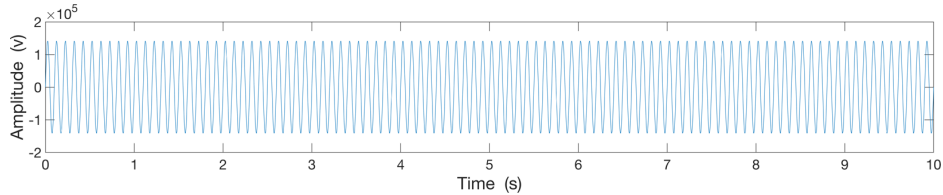

Figure 5. The electrical system line voltage 60Hz, 100kV rms value.

In order to enhance the visibility of the modulation by the DW noise $N_w(t)$, in Equation 3, we use a Gaussian noise with 0.3 rms value and bandwidth 1Hz, which are not practical but useful for illustration purposes, see Figure 6.

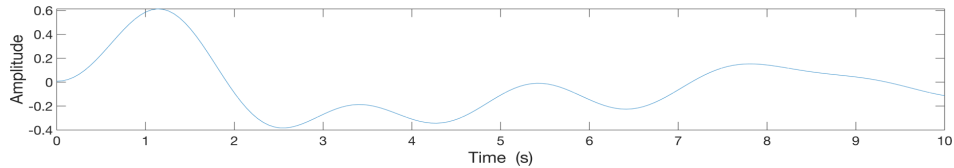

Figure 6. A DW noise $N_w(t)$ with bandwidth 1 Hz.

Similarly, for the parasitic noise $N_p(t)$ that is additive to $N_w(t)$, we use a Gaussian noise with 0.2 rms value and bandwidth 0.5Hz, see Figure 7.

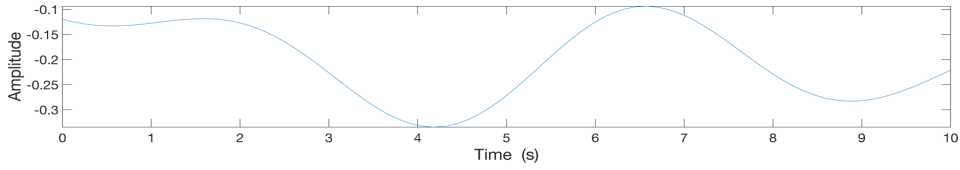

Figure 7. The parasitic noise $N_p(t)$ with bandwidth 0.5 Hz.

Figure 8 shows the sum of the two noises, which amplitude modulates the line voltage by the factor $[1+ N_w(t)+N_p(t)]$, cf. Equation 3, which is for the ideal case free of parasitic noise.

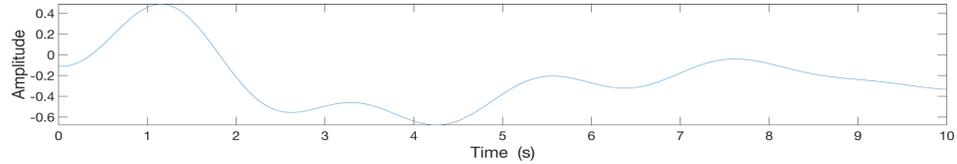

Figure 8. The sum of $N_w(t)$ and $N_p(t)$.

In Figure 9, the watermarked line voltage is shown (cf. Equation 5). Here the parasitic noise is also added to the watermarking noise to illustrate a practical situation, the amplitude of the modulating noises are shown greatly exaggerated to illustrate the concept.

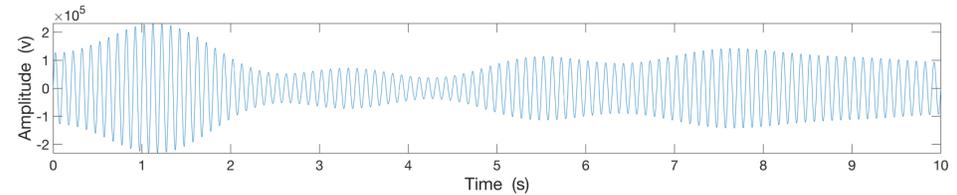

Figure 9. The line voltage modulated by the multiplicative factor $[1+N_w(t)+N_p(t)]$, cf. Equation 5. The amplitude of the noise has been greatly exaggerated to illustrate the concept.

Figure 10 shows the demodulated line voltage by the sensor before the low-pass filtering, see Equation 6. As it is already mentioned, the parasitic noise is also included.

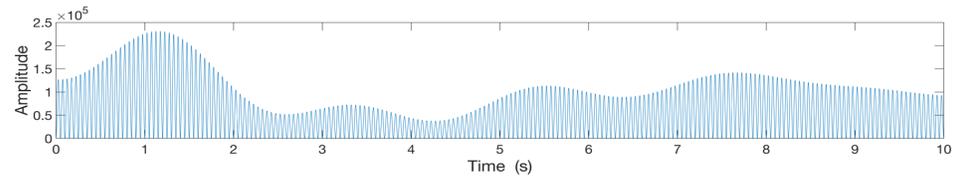

Figure 10. The demodulated line noise with the additive 120Hz second harmonic, the DC and the noise components before low-pass filtering (or short-time tome averaging) cf. Equation 6.

Figure 11 shows the regular output signal $[1+N_w(t)+N_p(t)]$ of the sensor after low-pass filtering and normalizing with the nominal peak voltage $a_0$, cf. Equation 7.

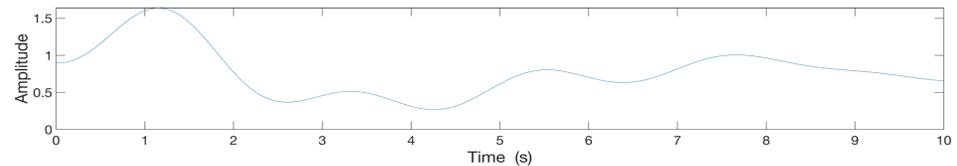

Figure 11. The sensor output $[1+N_w(t)+N_p(t)]$ after low-pass filtering, normalized by $a_0$, cf. Equation 7. The value 1 indicates 100 kV rms value.

After subtracting 1 from the sensor output data shown in Figure, Trudy obtains the sum of $N_w(t)$ and $N_p(t)$ that she can use to synthesize the fake sensor signal. In Figure 12, an attack is shown. In Equation 13, $\alpha = \beta = \gamma = 0.5$ is chosen to generate $S_f(t)$.

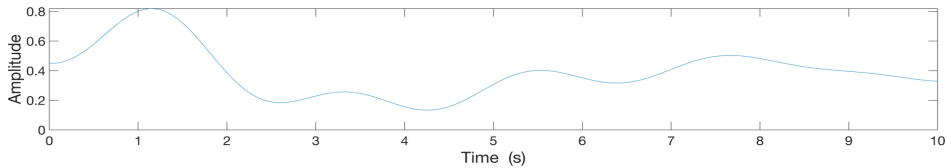

Figure 12. Example for the fake sensor signal $S_f(t)$ where both the nominal voltage and the noises are half of their actual values. It can imitate a transformer failure.

Conrad will see the DW noise there and interpret the situation as an equipment malfunction or failure that leads to a reduced grid voltage, such as a transformer fault instead of an attack.

**Conclusion**

In this study, we further investigated the security guarantees provided by dynamic watermarking in smart grid applications. The spectral analysis of the DW protocol revealed limitations in the allowable noise bandwidth of the watermarking noise. Building upon our previous findings that DW cannot offer unconditional security, we developed a simple yet effective attack. Similar to the previous attack involving a digital twin, the new attack does not require knowledge of the private watermarking signal used by the grid controller. Due to the Controller's inability to detect the ongoing breach, the attacker can fully expose the grid.

These results corroborate our earlier assessment that without secure or authenticated communication channels and tamper-resistant sensors, DW fails to provide any meaningful security assurances, neither conditional nor unconditional. Furthermore, we find that when communication links, sensors, and other components are made tamper-resistant through the use of secure and authenticated connections, the need for DW becomes redundant. The grid can be effectively secured through these fundamental security measures alone, rendering the additional complexity of dynamic watermarking unnecessary.

It is noteworthy that secure communications and tamper-resistant devices require secure key exchange. For unconditional security in the grid, the key exchange must also be unconditionally secure. This finding highlights the importance of implementing robust key exchange mechanisms to ensure the overall security of the smart grid system.

**References**

[1] K. Davis, L. B. Kish, C. Singh, Smart Grids Secured by Dynamic Watermarking: How Secure?, *Fluct. Noise Lett.* **23** (2024) 2450043 (10 pages), DOI: 10.1142/S0219477524500433 .
[2] B. Satchidanandan, P. R. Kumar, Dynamic Watermarking: Active Defense of Networked Cyber–Physical Systems, *Proc. IEEE* **105** (2017) 219-240.
[3] B. Satchidanandan, P. R. Kumar, On minimal tests of sensor veracity for dynamic watermarking-based defense of cyber-physical systems, 9th International Conference on Communication Systems and Networks (COMSNETS), Bengaluru, India, 2017, pp. 23-30, doi: 10.1109/COMSNETS.2017.7945354.
[4] T. Huang, J. Ramos-Ruiz, W.-H. Ko, J. Kim, P. Enjeti, P. R. Kumar, and L. Xie, Enabling Secure Peer-to-Peer Energy Transactions Through Dynamic Watermarking in Electric Distribution Grids, *IEEE Electrification Magazine* 9 (2011) 55-64. DOI: 10.1109/MELE.2021.3093600.
[5] T-H. Lin, P. R. Kumar, The Dynamic Watermarking Method for the Cybersecurity of the Tennessee Eastman Process Control System, 2023 Ninth Indian Control Conference (ICC), 18 - 20 December, 2023,

*IEEE*, DOI: 10.1109/ICC61519.2023.10441912
[6] Digital Twin for Power Equipment, White Paper, IEEE (January 2024), https://resourcecenter.ieee-pes.org/publications/white-papers/pes_tp_wp_dtwin_013024
[7] P. Kundur, Power system stability, in Power System Stability and Control, CRC Press 2012, New York.